\documentclass[runningheads]{llncs}
\usepackage{cite}
\usepackage{graphicx}
\usepackage{siunitx}
\usepackage[acronym]{glossaries}
\glsdisablehyper
\newacronym{gdp}{GDP}{Gross Domestic Product}

\newacronym{bim}{BIM}{Building Information Modeling}
\newacronym{aec}{AEC}{Architecture, Engineering, and Construction}
\newacronym{aec-fm}{AEC/FM}{Architecture, Engineering, Construction, and Facility Management}

\newacronym{ar}{AR}{Augmented Reality}
\newacronym{av}{AV}{Augmented Virtuality}
\newacronym{vr}{VR}{Virtual Reality}
\newacronym{mr}{MR}{Mixed Reality}
\newacronym{hmd}{HMD}{Head-Mounted Display}

\newacronym{iot}{IoT}{Internet of Things}

\usepackage{hyperref}
\usepackage{cleveref}
\usepackage{subfig}
\usepackage[table]{xcolor}
\DeclareUnicodeCharacter{25CF}{$\bullet$}
\begin{document}
\title{Supporting Construction and Architectural Visualization through BIM and AR/VR: A Systematic Literature Review}
\titlerunning{Construction and Architectural Visualization through BIM and AR/VR}
%
\author{Enes Yigitbas \and Alexander Nowosad \and Gregor Engels}

\authorrunning{E. Yigitbas et al.}

\institute{Paderborn University\\ Zukunftsmeile 2, 33102 Paderborn, Germany\\
\email{firstname.lastname@upb.de}, 
}
\maketitle              
\begin{abstract} 
  The \gls{aec-fm} industry deals with the design, construction, and operation of complex buildings.
  Today, \gls{bim} is used to represent information about a building in a single, non-redundant representation.
  Here, \gls{ar} and \gls{vr} can improve the visualization and interaction with the resulting model by
  augmenting the real world with information from the \gls{bim} model or allowing a user to immerse in a virtual world generated from the \gls{bim} model. This can improve the design, construction, and operation of buildings. While an increasing number of studies in HCI, construction, or engineering have shown the potential of using AR and VR technology together with BIM, often research remains focused on individual explorations and key design strategies. In addition to that, a systematic overview and discussion of recent works combining AR/VR with BIM are not yet fully covered. Therefore, this paper systematically reviews recent approaches combining \gls{ar}/\gls{vr} with \gls{bim} and categorizes the literature by the building's lifecycle phase while systematically describing relevant use cases. In total, 32 out of 447 papers between 2017 and 2022 were categorized. The categorization shows that most approaches focus on the construction phase and the use case of review and quality assurance. In the design phase, most approaches use VR, while in the construction and operation phases, AR is prevalent. 

\end{abstract}
\glsresetall

\section{Introduction}
\label{sec:introduction}

The \gls{aec} industry is essential for a country's economy \cite{pheng2019economy}. For example, in Germany in 2021, the investments into construction were 416.700 billion euros, 11:6\% of the Gross Domestic Product (GDP) \cite{destatis2022vgr}.
Construction projects are complex, especially
in organizational and technological aspects \cite{luo2017construction}. Thus, stakeholders in the industry need to
be supported with tools to increase their performance. In the past, information about the
project has been separated into different paper documents, CAD models, and other tools, leading to inconsistencies
and inefficient communication \cite{eastman2008bim}.

Here, \gls{bim} is a technology that avoids this fragmentation
and prevents inconsistencies by combining the information into a non-redundant representation that
can be viewed from different views \cite{eastman2008bim}. This supports the entire lifecycle of a building, from the
design phase over the construction up to the operation and maintenance leading to improvements for the
\gls{aec-fm} industry \cite{eastman2008bim}.
The advantages of \gls{bim} are, for example, precise visualization of
a building in the design phase and simulation of the construction process by adding time as a fourth dimension to the
model \cite{eastman2008bim}. Currently, printed 2D plans, computers, and tablets are used to show
2D or 3D models of a building.
This brings many disadvantages and challenges. For example, a customer cannot see the real sizes of geometries on a computer screen, leading to
wrong design decisions.
Furthermore, a project manager has to manually map aspects displayed on a computer to structures in the real world
to check the progress of a building, which is time-consuming.
A facility manager has to orient via a 2D plan of a building and use a 2D plan of assets to repair them.

\Gls{vr} and \gls{ar} have been a topic of intense research in the last decades \cite{doerner2022virutal, carmigniani2011augmented}. In the past few years, massive advances in affordable consumer hardware and accessible software frameworks are now bringing these technologies to the masses. While \Gls{vr} is a technology that allows a user to immerse in a virtual world, \gls{ar} adds virtual objects to the real world \cite{milgram1994taxonomy}. These technologies can overcome the visualization issues of \gls{bim} models in different phases of the building's lifecycle. For example, a customer could experience different designs of a building through \gls{vr} in an immersive way \cite{rahimian2019openbim, anthes2016state},
a project manager can use \gls{ar} to check the progress of a building and the building structures are augmented with information
about their progress \cite{meza2014component, ratajczak2019bim}. A facility manager
can leverage \gls{ar} to find failed assets in a building and get step-by-step guides on how to repair the assets \cite{xie2020visualized, chew2020evaluating}.

While an increasing number of such approaches and studies in HCI, construction, or engineering have shown the potential of using AR and VR technology together with BIM, often research remains focused on individual explorations and key design strategies. In addition to that, a systematic overview and discussion of recent works combining AR/VR with BIM are not yet fully covered. Therefore, this paper investigates recent developments in the field of using \gls{ar} and \gls{vr} as visualization tools for \gls{bim} models
in all phases of a building's lifecycle with a focus on use cases and answers the following research questions (RQs):

\textit{RQ1: What are the current use cases for using \gls{ar} and \gls{vr} together with \gls{bim} in the \gls{aec-fm} industry?}

\textit{RQ2: For which use cases is \gls{ar} and for which use cases is \gls{vr} used and why?}

To answer the research questions, we have conducted a systematic literature review where we review recent approaches combining \gls{ar}/\gls{vr} with \gls{bim}. The main goal of this literature review is to categorize the literature by the building's lifecycle phase while systematically describing relevant use cases. 

The remainder of this paper is as follows. \Cref{sec:background} gives background information about important aspects of
the research topic. \Cref{sec:relatedWork} shows other surveys that have already dealt with the topic. \Cref{sec:methodology} describes the
methodology used to gather literature. \Cref{sec:results} presents the results from the literature review. \Cref{sec:furtherResearch}
sketches gaps in the literature that have the potential for further research. \Cref{sec:limitations} shows the limitations
of this work. Finally, \Cref{sec:conclusion} summarizes this paper and shows potential for future work.

\section{Background}
\label{sec:background}

In this section, we briefly describe essential concepts from the \gls{aec-fm} industry focusing on the lifecycle phases of a building and the \gls{bim}.

\subsection{Lifecycle phases of a building}
\label{sec:background:phases}

The lifecycle phases of a building are essential to categorize applications based on the building's lifecycle phase in which they are used.
The lifecycle of a building can be broken down into different phases. Multiple options exist, e.g., Eadie et al. \cite{eadie2013bim} split the field
into feasibility, design, preconstruction, construction, and operation and management, while Arditi and Gunaydin \cite{arditi1997total} use the three phases of
planning and design, construction, and maintenance and operation. This paper uses the phases of \emph{Design}, \emph{Construction}, and \emph{Operation}.

Here, the \emph{Design} phase includes the feasibility and actual design of a building. Feasibility is the initial planning of a building by defining the main
goals, a broad schedule, and a budget \cite{dykstra2018construction}. Then, the actual design of a building is done by creating detailed construction documents
that define the building \cite{dykstra2018construction}.

The \emph{Construction} phase includes all tasks necessary for the construction, which are pre-construction and actual construction. Pre-construction
is the organization of the construction works \cite{dykstra2018construction}. This includes detailed planning of the construction works and planning of the
construction site layout, e.g., where machines and materials are placed \cite{dykstra2018construction}. After pre-construction, the actual construction works
start where construction workers are constructing the building \cite{dykstra2018construction}.

The \emph{Operation} phase includes all maintenance and operation tasks after the building is finished. Here, this is taking care of the repairing of
parts of the building to retain the expected functionality \cite{allen1993building}. This can also include monitoring the building via sensor values
\cite{xie2020visualized}.

\subsection{BIM}
\glsreset{bim}
\Gls{bim} technology allows the precise construction of a digital virtual model of a building \cite{eastman2008bim}.
Here, this model can be viewed in different views, for example, as a 2D plan or 3D model, that are all consistent with each other \cite{eastman2008bim}.
Furthermore, \gls{bim} is not only the model but also has linked processes to create, communicate, and evaluate this model \cite{eastman2008bim}.
These processes are supported by the model, which itself is not only 3D geometric data
but consists of so-called parametric objects that contain the geometry and are associated with additional data and rules \cite{eastman2008bim}. Additionally, it is
possible to add behavioral data to objects to allow evaluations and simulations \cite{eastman2008bim}. Here, consistency plays an important role
and is achieved by the non-redundant representation of geometries in the \gls{bim} model \cite{eastman2008bim}.
Moreover, if a parametric object in a \gls{bim} model is changed, all related objects are updated, too \cite{eastman2008bim}.

Through this, \gls{bim} technology can enhance the whole lifecycle of a building \cite{eastman2008bim}.
In the design phase, there is always one consistent representation of the building \cite{eastman2008bim}. This representation allows for early
visualizations in 3D, and the exports, e.g., for 2D plans, in any stage are always consistent \cite{eastman2008bim}. Additionally,
design changes are automatically reflected in all building parts \cite{eastman2008bim}. The \gls{bim} model can also be used
for early automatic evaluations, for example, to check energy efficiency \cite{eastman2008bim}.

In the construction phase, \gls{bim} allows for the simulation of the construction process by associating a construction plan with the
parametric objects in the \gls{bim} model \cite{eastman2008bim}. This simulation can reveal on-site conflicts and safety issues \cite{eastman2008bim}.
Additionally, it supports scheduling material orders and deliveries \cite{azhar2011building}. Furthermore,
design changes in this phase are automatically updated in the \gls{bim} model, and all consequences of the change are visible \cite{eastman2008bim}.
Finally, the operation phase is supported by the precise and up-to-date \gls{bim} model through the knowledge of all spaces and systems in the building \cite{eastman2008bim}.

\section{Related Work}
\label{sec:relatedWork}

Augmented Reality (AR) and Virtual Reality (VR) have been a topic of intense research in the last decades. While VR interfaces support the interaction in an immersive computer-generated 3D world and have been used in different application domains such as training~\cite{DBLP:conf/vrst/YigitbasJSE20}, education~\cite{DBLP:conf/mc/YigitbasTE20}, modeling~\cite{DBLP:conf/models/YigitbasGWE21}, prototyping \cite{DBLP:conf/vl/YigitbasKGE21}, or evaluation~\cite{DBLP:conf/hcse/KarakayaYE22}, AR enables the augmentation of real-world physical objects with virtual elements and has been also applied in various application domains such as product configuration (e.g.,\cite{DBLP:conf/hcse/GottschalkYSE20}, \cite{DBLP:conf/hcse/GottschalkYSE20a}), prototyping \cite{DBLP:conf/hcse/JovanovikjY0E20}, planning and measurements~\cite{DBLP:conf/eics/EnesScaffolding}, robot programming (e.g., \cite{DBLP:conf/eics/KringsYBE22}, \cite{DBLP:conf/interact/YigitbasJE21}), or for realizing smart interfaces (e.g., \cite{DBLP:conf/eics/KringsYJ0E20}, \cite{DBLP:conf/interact/YigitbasJ0E19}).

In the following, we especially focus on and present other surveys that deal with the combination of \gls{bim} and \gls{ar} or \gls{vr}.
Additionally, we describe what makes this paper different from the related work and why a novel and updated literature review is required on this topic.

Calderon-Hernandez and Brioso \cite{hernandez2018lean} survey papers that use
\gls{ar} and \gls{bim} together in the design and construction phases.
Here, they focus on the last five years from 2018 backward and main journals in the field of construction planning neglecting HCI, especially AR/VR relevant venues. 

Wu et al. \cite{wu2021integrated} conduct an application-focused literature review. They show gaps, challenges,
and future work in the field. Furthermore, they classify the papers by their application category. The application categories,
which they defined, are task guidance and information retrieval, design and refinement, process planning and control, and
upskilling of the \gls{aec} workforce.

Sidani et al. \cite{sidani2021tools} perform a survey of \gls{bim}-based \gls{vr}. They classify the approaches
by the research field, the \gls{bim} dimensions, construction stages, and target groups. Here, the research field
can be collaboration, construction design, construction management, construction safety, education, or facility management.

Salem et al. \cite{salemBIM2020} show different use cases for \gls{ar} and \gls{vr}. They state that \gls{vr} is
used for visualization and \gls{ar} in broader applications. However, they do not provide a classification and also do not
have a concrete methodology.

Wang et al. \cite{wang2013augmented} develop a classification framework for \gls{ar} in the built environment.
Their classification framework consists of multiple parts. One of them categorizes the application area. They come up with
the following areas: architecture and design, landscape and urban planning, engineering-related, construction, facility management,
life cycle integration, and training and education. The literature that is reviewed is from 2005 to 2011.

Compared to the above-mentioned works, this paper considers both, \gls{ar} and \gls{vr} approaches combined with BIM. Additionally, all building's lifecycle phases are considered, and the surveyed literature is categorized by the building's lifecycle phase and the use case category. With this regard, our work contributes a systematic and holistic overview of recent approaches where AR and VR technologies are combined with BIM.   

\section{Methodology}
\label{sec:methodology}

In this section, we describe the methodology of the research.

\begin{figure}[h]
  \includegraphics[width=0.9\textwidth]{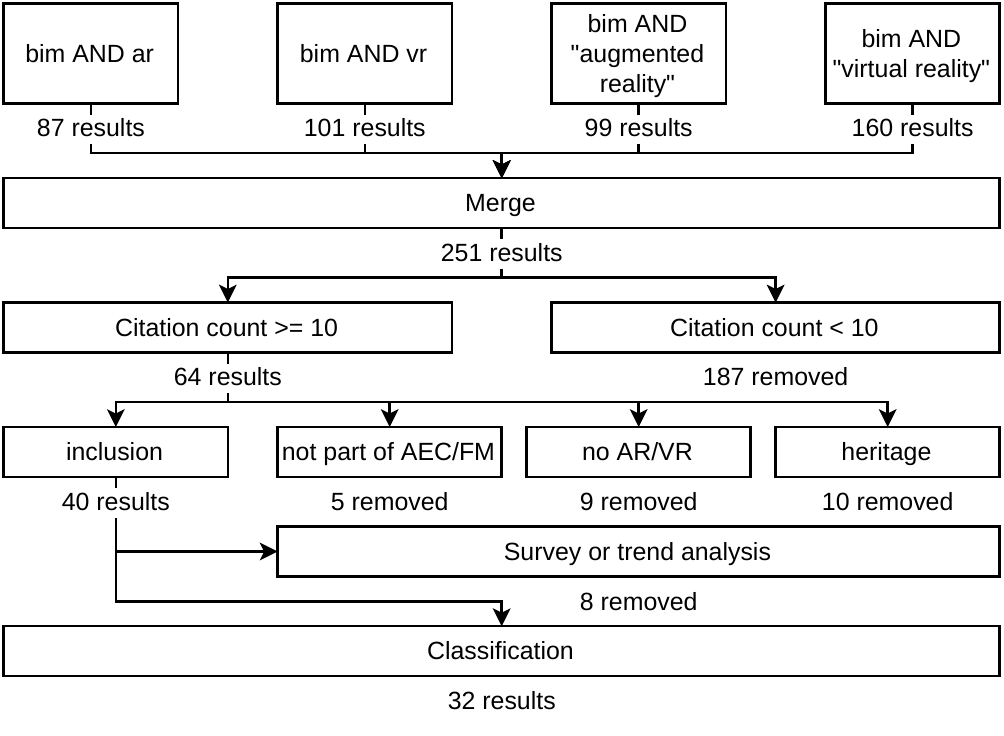}
  \caption{Methodology of the conducted systematic literature review}
  \label{figure:literature-review}
\end{figure}

To answer the research questions, which are defined in \Cref{sec:introduction},
we conducted a systematic literature review based on \cite{xiao2019guidance} focusing on concrete use cases in the
\gls{aec-fm} industry that combine \gls{ar}/\gls{vr} with \gls{bim}.
An overview of the review methodology is shown in \Cref{figure:literature-review}. To find literature,
we used Scopus\footnote{www.scopus.com}. Here, we used the following search terms:
\emph{bim AND ar} (87 results), \emph{bim AND vr} (101 results),
\emph{bim AND "augmented reality"} (99 results), and
\emph{bim AND "virtual reality"} (160 results). For all search terms, the search fields were set to
article title, abstract, and keywords. Additionally, the search was limited to open-access
articles. Furthermore, only articles between 2017 and 2022 were considered, the language was set to
English, and the document type was restricted to article or conference paper.

After exporting the results from Scopus, duplicates were removed, resulting in 251 unique results.
Then, papers that were less than ten times cited according to Scopus were removed to only consider established and impactful papers with a certain relevance,
resulting in 64 remaining papers. After that, irrelevant papers were removed. Here, five were removed because they
are not part of the \gls{aec-fm} industry, and nine were removed because they use neither \gls{ar} nor \gls{vr}. Ten were removed
because they focused on heritage and not on the design, construction, or operation of current buildings.
Eight of the remaining 40 papers were removed because they are surveys or trend analyses and do not present a
concrete use case that could be analyzed. After these filtering steps, 32 papers remain.

The remaining papers were categorized by whether they use \gls{ar} or \gls{vr} or both,
which device they use, in which building's lifecycle phase the approach they present is used,
and whether they present a prototype or only sketch the solution. Additionally, the use case
they present was categorized.

\section{Results}
\label{sec:results}

In this section, we present the results of the literature review. An overview of the categorization
of the 32 papers is shown in \Cref{table:approaches}.

{
\begin{table}
\rowcolors{2}{gray!35}{}
\begin{tabular}{|l|c|cc|c|ccc|c|ccccc|c|}
\hline
&  &  &  &  & \multicolumn{3}{c|}{ Phase } &  & \multicolumn{5}{c|}{ Use Case } &  \\
\hline
  Publication                       & Year & \rotatebox{90}{AR} & \rotatebox{90}{VR} & \rotatebox{90}{Device}     & \rotatebox{90}{Design} & \rotatebox{90}{Construction} & \rotatebox{90}{Operation} & \rotatebox{90}{Prototype} & \rotatebox{90}{Planning} & \rotatebox{90}{Review \& Quality Assurance } & \rotatebox{90}{Task guidance} & \rotatebox{90}{Safety} & \rotatebox{90}{Education} & \rotatebox{90}{Localization \& Tracking} \\
\hline
  Afzal, M. et al.  \cite{afzal2021evaluating}             & 2021 &    & ●  & HMD        &        & ●            &           & ●         &          &                            &               & ●      & ●         &                         \\
  Azhar, S. \cite{azhar2017role}                     & 2017 &    & ●  & HMD        &        & ●            &           & ●         &          &                            &               & ●      & ●         &                         \\
Chalhoub, J. and Ayer, S. K. \cite{chalhoub2018mixed} & 2018 & ●  &    & HMD        &        & ●            &           & ●         &          &                            & ●             &        &           &                         \\
Chalhoub, J. et al. \cite{chalhoub2021augmented}          & 2021 & ●  &    & HMD        &        & ●            &           & ●         &          & ●                          &               &        &           &                         \\
Chew, M. Y. L. et al. \cite{chew2020evaluating}        & 2020 & ●  & ●  & -          &        &              & ●         &           &          &                            & ●             &        & ●         &                         \\
Dallasega, P. et al. \cite{dallasega2020bim}         & 2020 & ●  & ●  & HMD        &        & ●            &           & ●         & ●        &                            & ●             &        &           &                         \\
Diao, P.-H. and Shih, N.-J. \cite{diao2019bim}  & 2019 & ●  &    & Mobile     &        &              & ●         & ●         &          &                            & ●             & ●      &           &                         \\
Garbett, J. et al. \cite{garbett2021multiuser}           & 2021 & ●  &    & Mobile     & ●      & ●            &           & ●         & ●        &                            &               &        &           &                         \\
Gomez-Jauregui, V. et al. \cite{gomez2019quantitative}    & 2019 & ●  &    & Mobile     &        & ●            &           & ●         &          &                            &               &        &           & ●                       \\
Hasan, S. M. et al. \cite{hasan2022augmented}          & 2022 & ●  &    & Mobile     &        & ●            &           & ●         &          &                            &               & ●      &           &                         \\
Herbers, P. and König, M. \cite{herbers2019indoor}    & 2019 & ●  &    & HMD        &        & ●            & ●         & ●         &          &                            &               &        &           & ●                       \\
Hernández, J. L. et al. \cite{hernandez2018ifc}      & 2018 & ●  &    & Mobile     &        & ●            &           & ●         &          & ●                          &               &        &           &                         \\
Hübner, P. et al. \cite{huebner2018marker}            & 2018 & ●  &    & HMD        &        & ●            &           & ●         &          &                            &               &        &           & ●                       \\
Kamari, A. et al. \cite{kamari2021bim}            & 2021 &    & ●  & M.-HMD     & ●      &              &           & ●         &          & ●                          &               &        &           &                         \\
Khalek, I. A. et al. \cite{khalek2019augmented}         & 2019 & ●  &    & HMD        & ●      &              &           & ●         &          & ●                          &               &        &           &                         \\
Lou, J. et al. \cite{lou2017study}               & 2017 & ●  &    & -          &        & ●            &           &           &          & ●                          &               &        &           &                         \\
Mahmood, B. et al. \cite{mahmood2020bim}           & 2020 & ●  &    & HMD        &        & ●            &           & ●         &          &                            &               &        &           & ●                       \\
Mirshokraei, M. et al.\cite{mirshokraei2019web}       & 2019 & ●  &    & Mobile     &        & ●            &           & ●         &          & ●                          &               &        &           &                         \\
Natephra, W. et al. \cite{natephra2017integrating}          & 2017 &    & ●  & HMD        & ●      &              &           & ●         & ●        &                            &               &        &           &                         \\
Pour Rahimian, F. et al. \cite{rahimian2019openbim}     & 2019 &    & ●  & HMD        & ●      &              &           & ●         & ●        &                            &               &        &           &                         \\
Pour Rahimian, F. et al. \cite{rahimian2020ondemand}     & 2020 &    & ●  & HMD        &        & ●            &           & ●         &          & ●                          &               &        &           &                         \\
Ratajczak, J. et al. \cite{ratajczak2019bim}         & 2019 & ●  &    & Mobile     &        & ●            &           & ●         &          & ●                          &               &        &           &                         \\
Riexinger, G. et al. \cite{riexinger2018mixed}         & 2018 & ●  &    & Mobile\&HMD &        & ●            &           & ●         &          & ●                          & ●             &        &           &                         \\
Saar, C. C. et al. \cite{saar2019bim}           & 2019 & ●  &    & Mobile     &        & ●            & ●         & ●         &          & ●                          & ●             &        & ●         &                         \\
Schranz, C. et al. \cite{schranz2021potentials}           & 2021 & ●  &    & Mobile\&HMD & ●      &              &           &           &          & ●                          &               &        &           &                         \\
Schweigkofler, A. et al. \cite{schweigkofler2018development}     & 2018 & ●  &    & Mobile     &        & ●            &           & ●         &          & ●                          & ●             &        &           &                         \\
Vasilevski, N. and Birt, J. \cite{vasilevski2020analysing}  & 2020 &    & ●  & M.-HMD     & ●      &              &           & ●         &          &                            &               &        & ●         &                         \\
Ventura, S. M. et al. \cite{ventura2020design}        & 2020 &    & ●  & Other      & ●      &              &           & ●         &          & ●                          &               &        &           &                         \\
Vincke, S. et al. \cite{vincke2019immersive}            & 2019 &    & ●  & HMD        &        & ●            &           & ●         &          & ●                          &               &        &           &                         \\
Xie, X. et al. \cite{xie2020visualized}               & 2020 & ●  &    & HMD        &        &              & ●         & ●         &          &                            & ●             &        &           &                         \\
Zaher, M. et al. \cite{zaher2018mobile}             & 2018 & ●  &    & Mobile     &        & ●            &           & ●         &          & ●                          &               &        &           &                         \\
Zaker, R. and Coloma, E. \cite{zaker2018virtual}     & 2018 &    & ●  & HMD        & ●      &              &           & ●         &          & ●                          &               &        &           &                         \\
\hline
\end{tabular}
\caption{Categorization of identified approaches related to AR/VR and BIM}
\label{table:approaches}
\end{table}
}

The approaches are categorized by whether they use \gls{ar} or \gls{vr}. For \gls{ar}, the definition of Azuma \cite{azuma1997survey} is used. Here, already the navigation in the \gls{ar} environment is counted as real-time interactivity. For \gls{vr},
fully and partly immersive solutions are considered. This means it is enough for a solution to use a 3D screen.
If an approach uses \gls{ar} and \gls{vr}, both cells are marked. It is also possible that an approach uses \gls{ar} and \gls{vr} for
different use cases. These cases are not visible in the table but in the detailed descriptions below.

The devices are classified into different groups.
The group \emph{HMD} contains all consumer
\glspl{hmd} developed for \gls{ar} or \gls{vr}, e.g.,
a Microsoft HoloLens\footnote{\url{https://www.microsoft.com/en-us/hololens}} for \gls{ar}
or a Valve Index\footnote{\url{https://store.steampowered.com/valveindex}} for \gls{vr}.
Further, \mbox{\emph{M.-HMD}} stands for mobile \gls{hmd} and contains low-cost devices where a smartphone is mounted as an \gls{hmd}, e.g.,
Google Cardboard\footnote{\url{https://arvr.google.com/cardboard/}}.
\emph{Mobile} contains all handheld devices, like smartphones or tablets, and \emph{Other} contains
all devices that do not fit into this classification. Additionally, if an approach uses multiple devices,
they are connected by an and (\&), and if they are using different devices for \gls{ar} and \gls{vr}, the
device for \gls{ar} is separated from the device for \gls{vr} via a slash (/).

The building's lifecycle phases used for the categorization are \emph{Design}, \emph{Construction}, and
\emph{Operation}, as described in \Cref{sec:background:phases}. If an approach is used in multiple phases, both are marked.

If a paper also presents a prototype to show the approach, the mark in the \emph{Prototype} column is set. Otherwise,
the approach is only described in the paper. If an approach is only described and does not present a prototype, the
device column might be empty if a concrete device type is not mentioned.

The use cases are grouped into \emph{Planning}, \emph{Review \& Quality Assurance}, \emph{Task guidance},
\emph{Safety}, and \emph{Education}. A detailed description of these use case categories is given in \Cref{sec:results:useCases}.
If an approach supports multiple use cases, all categories that can be applied are marked. Here, the table does not show which use cases are used in
which building's lifecycle phases if multiple use case categories are marked. This interrelation is only visible in the detailed description below.
Additionally, there is the category \emph{Localization \& Tracking} for approaches that do not focus on concrete use cases but on localization
and tracking improvements on-site.

General statistics about the reviewed approaches show that \gls{ar} is used by most approaches with \SI{69}{\percent}, and \gls{vr} is only used by \SI{31}{\percent}
of the approaches. Over half of the approaches use an \gls{hmd} as a device to achieve \gls{vr} or \gls{ar}, followed by
mobile devices with \SI{38}{\percent}. Only two approaches use a mobile \gls{hmd}, and only one approach uses another
device, in this case, a special 3D screen \cite{ventura2020design}.

The prevalent phase in which \gls{ar} or \gls{vr} are used with \gls{bim} is the \emph{Construction} phase, with
\SI{56}{\percent} of the approaches. Only \SI{28}{\percent} of the approaches are used in the \emph{Design} phase and
\SI{16}{\percent} in the \emph{Operation} phase. Almost all papers present a prototype of their approach.

The remainder of this section is as follows. \Cref{sec:results:useCases} focuses on the first research question by
defining a classification for the use cases and describing them. \Cref{sec:results:arvr} deals with the second research
question by showing when \gls{ar} is preferred and when \gls{vr} is preferred and giving reasons for this.
\Cref{sec:results:discussion} sums up the results to answer the research questions.

\subsection{Use Cases for Combining BIM with AR/VR}
\label{sec:results:useCases}

Based on the elicited literature, five use case categories for combining \gls{ar} or \gls{vr}
with \gls{bim} are defined. The use case categories are independent of the building's lifecycle
phase in which they are used.

The first category is \emph{Planning}. It is given whenever the \gls{bim} model is used for planning purposes, e.g.,
to plan the design of a building, to plan the construction, or to plan maintenance tasks.
The second category is \emph{Review \& Quality Assurance}. It includes all use cases where the \gls{bim} model is used for reviewing tasks, e.g.,
to review the design represented as a \gls{bim} model, to review the construction works based on the planning in a \gls{bim} model, or to review maintenance tasks.
\emph{Task guidance} groups use cases that help a user to complete a task, e.g., by giving them information about the task, concrete steps
to complete it, or visualizing the task or asset to work on. The \emph{Safety} use case is given if
an approach increases the safety of workers or facility managers on-site. The last use case, \emph{Education},
includes all approaches that can be used to educate stakeholders prior to the actual design, on-site work, or maintenance.

Additionally, there is the category \emph{Localization \& Tracking}, which is not directly a use case, but groups approaches that
focus on the technical aspects of \gls{ar} on-site. These approaches do not present concrete use cases
but based on their findings, multiple other use cases on-site become feasible. They make them feasible by allowing the localization of the user
on-site and the improvement of the tracking of the device to allow precise positioning of virtual objects in an \gls{ar}
environment.

In the following, the different use cases in the different building's lifecycle phases are described.

\subsubsection{Design}
Nine approaches reside in the design phase
\cite{garbett2021multiuser, kamari2021bim, khalek2019augmented, natephra2017integrating,
rahimian2019openbim, schranz2021potentials, vasilevski2020analysing, ventura2020design, zaker2018virtual}.
Most of them either focus on the \emph{Planning} or the \emph{Review \& Quality Assurance} use case.

\begin{figure}
  \centering
  \subfloat[Interior design \cite{rahimian2019openbim}]{\label{fig:design:planning}\includegraphics[width=0.48\textwidth]{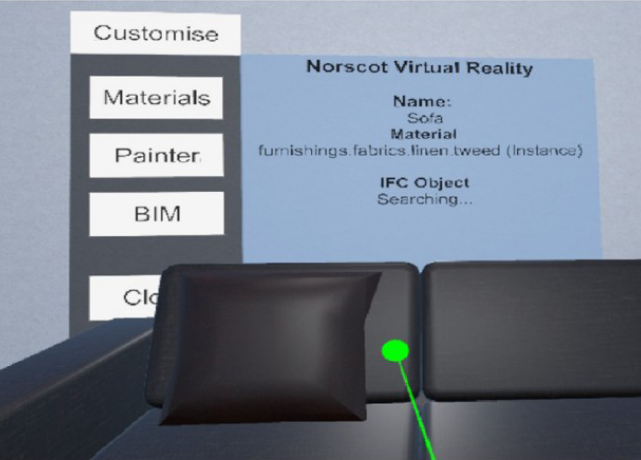}}
  \hfill
  \subfloat[Placement in the real world \cite{schranz2021potentials}]{\label{fig:design:ar}\includegraphics[width=0.48\textwidth]{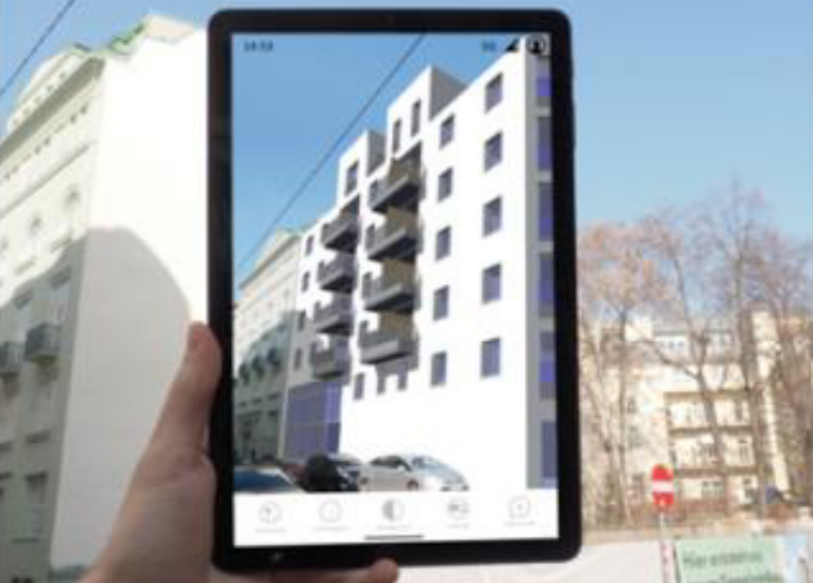}}%
  \caption{Design-related approaches in the area of AR/VR and BIM}
\end{figure}

For the \emph{Planning} use case category in the design phase \cite{garbett2021multiuser, natephra2017integrating, rahimian2019openbim}, the user
is provided with tools to edit the \gls{bim} model of a building in an \gls{ar}/\gls{vr} environment. The simplest realization
of this is a system where the users can collaboratively add annotations to the \gls{bim} model \cite{garbett2021multiuser}.
The two other approaches focus on interior design. Here, one approach \cite{rahimian2019openbim} shows the possibility of designing the painting of the walls
and allows the customization of the furniture. This approach is shown in \Cref{fig:design:planning}. Here, the user has the possibility to
change the material and color of a sofa \cite{rahimian2019openbim}.
The second approach \cite{natephra2017integrating} deals with the more specialized topic of
indoor lighting design. It allows the user to add different light bulbs to the room and simulate the resulting lighting situation.

The \emph{Review \& Quality Assurance} use case category in the design phase \cite{kamari2021bim, khalek2019augmented, schranz2021potentials, ventura2020design, zaker2018virtual}
allows the user to view a \gls{bim} model in an \gls{ar}/\gls{vr} environment to review the design.
Here, the focus can be on general architecture review \cite{zaker2018virtual} and can also be structured through a concrete protocol \cite{ventura2020design}.
One approach \cite{schranz2021potentials} describes the potential of this architecture review to be used for the regulatory process of checking a building
for compliance with legal rules. Here, instead of submitting printed plans, the \gls{bim} model is submitted and can be viewed in an \gls{ar}
environment by the building authority to check the model for compliance. This is shown in \Cref{fig:design:ar}. Here, a user can see the
building next to the surrounding buildings to check whether it fits into the environment \cite{schranz2021potentials}.

It is also possible to focus on more specialized topics. Here, one approach \cite{kamari2021bim} deals with the sustainability of a building's façade design by showing
the user different design alternatives in a \gls{vr} environment and letting them choose one based on the costs and sustainability. In this approach,
the virtual environment is only used to show the design to the user. The costs and sustainability are shown in an analog form. The \gls{bim} model is used to
generate the \gls{vr} environment and also to calculate the costs and sustainability. Two other approaches \cite{khalek2019augmented, zaker2018virtual}
use an \gls{ar} respectively a \gls{vr} environment to let the user test a design for maintainability. This testing is done by doing the maintenance steps
in the \gls{ar}/\gls{vr} environment and thus finding flaws in the design that hinder maintenance.

Only one approach \cite{vasilevski2020analysing} resides in the \emph{Education} use case category in the design phase. Here, a \gls{bim} model is explored in
a \gls{vr} environment to teach students. The approach allows the user to change the time in the simulation to see the model at different daytimes. Additionally, the approach
supports collaboration with other students to explore the model collaboratively.

The use case categories \emph{Task guidance} and \emph{Safety} are not used in the Design phase.

\subsubsection{Construction}

There are 21 approaches that reside in the construction phase
\cite{afzal2021evaluating, azhar2017role, chalhoub2018mixed, chalhoub2021augmented, dallasega2020bim,
garbett2021multiuser, gomez2019quantitative, hasan2022augmented, herbers2019indoor, hernandez2018ifc, huebner2018marker, lou2017study,
mahmood2020bim, mirshokraei2019web, rahimian2020ondemand, ratajczak2019bim, riexinger2018mixed, saar2019bim,
schweigkofler2018development, vincke2019immersive, zaher2018mobile}.
Over half of them focus on the \emph{Review \& Quality Assurance} use case category.

In the construction phase, the \emph{Planning} use case category deals with the actual planning of the construction works
\cite{dallasega2020bim, garbett2021multiuser}. Here, this can be viewing the \gls{bim} model
in \gls{vr} prior to construction to get an overview of the necessary construction steps \cite{dallasega2020bim} or
the possibility of annotating the \gls{bim} model off-site \cite{garbett2021multiuser}.

The \emph{Review \& Quality Assurance} use case category in the construction phase
\cite{chalhoub2021augmented, hernandez2018ifc, lou2017study,
mirshokraei2019web, rahimian2020ondemand, ratajczak2019bim, riexinger2018mixed, saar2019bim,
schweigkofler2018development, vincke2019immersive, zaher2018mobile} allows the user to check
the construction progress and identify potential flaws. Lou et al. \cite{lou2017study} describe different
potentials for using \gls{ar} for on-site quality management.
The potentials presented by them are the coordination through the information provided by the new applications, the possibility
to adapt the \gls{bim} model on-site to collect data and to account for rapid changes, and construction quality inspection.

One approach to support quality management is an information management system \cite{schweigkofler2018development}. Here,
users can select different objects on-site via an \gls{ar} interface and retrieve information, e.g., quality checklists \cite{schweigkofler2018development}. Additionally,
they can add information to the objects \cite{schweigkofler2018development}. It is also possible to build a specialized system for quality assurance that
allows users to use an \gls{ar} app on-site to go through the checklists and upload the inspection results to the
\gls{bim} model \cite{mirshokraei2019web}.

\begin{figure}
  \subfloat[Progress information \cite{ratajczak2019bim}]{\label{fig:construction:qa:ar:details}\includegraphics[width=0.48\textwidth]{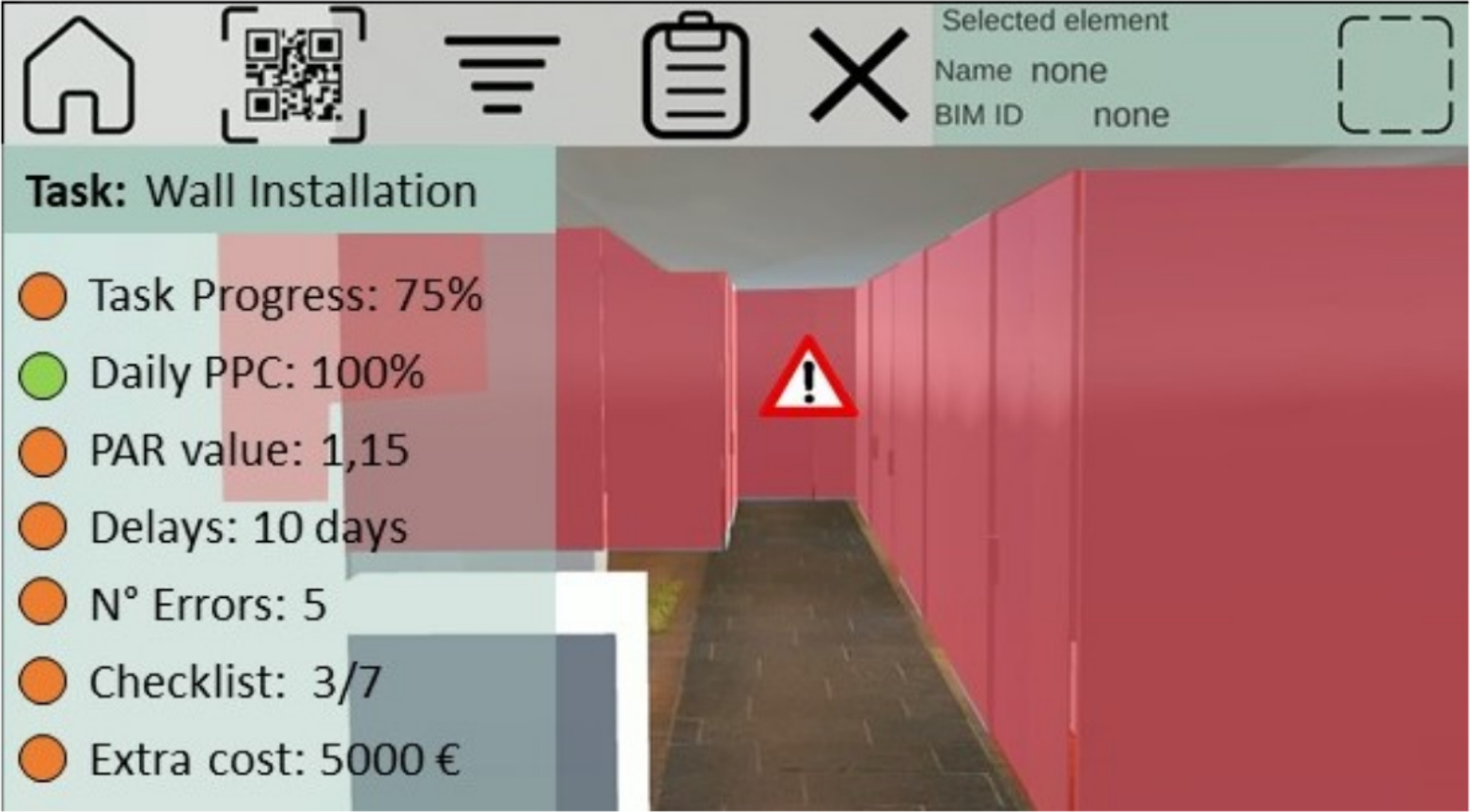}}
  \hfill
  \subfloat[Missing element \cite{riexinger2018mixed}]{\label{fig:construction:qa:ar:missing}\includegraphics[width=0.48\textwidth]{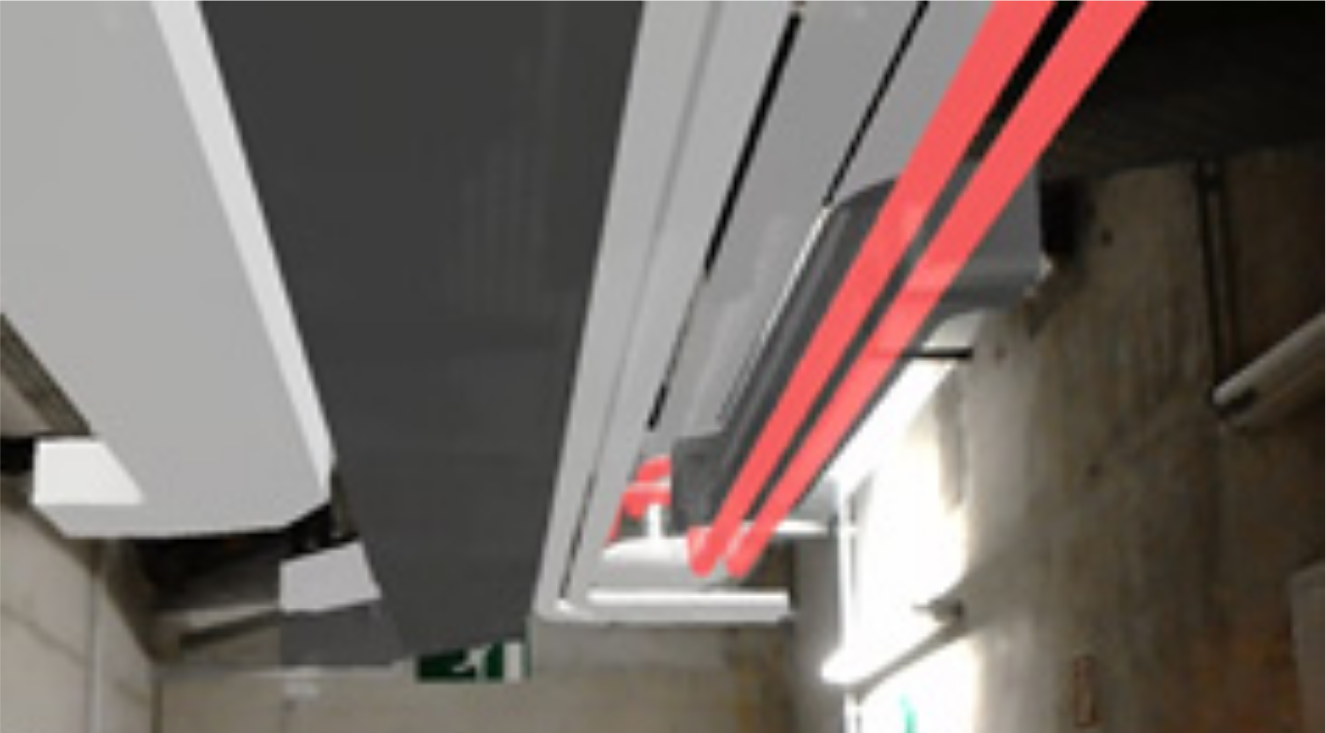}}%
  \caption{\Gls{ar}-related approaches that amend the real world with BIM information}
  \label{fig:construction:qa:ar}
\end{figure}

Additionally, such a system can be extended to show the progress information
by adding colored virtual elements to the real world and amending them with progress, costs, time, and quality information \cite{ratajczak2019bim}. In this approach,
site managers do not have to study 2D plans and other artifacts in their offices but can directly walk through the building under construction
and check the progress, quality, and next steps \cite{ratajczak2019bim}. The approach is shown in \Cref{fig:construction:qa:ar:details}. Here, different progress information about the highlighted wall structure
is visualized \cite{ratajczak2019bim}.

Another approach is to give this possibility directly to the workers to allow them to self-inspect their
work \cite{riexinger2018mixed, saar2019bim, hernandez2018ifc}. In these approaches, the real world is compared with the as-planned \gls{bim} model in \gls{ar}.
This comparison allows users to find deviations \cite{chalhoub2021augmented}.
\Cref{fig:construction:qa:ar:missing} shows an example of how a missing element in the real world could be visualized via an \gls{ar} virtual object.
Furthermore, there is the possibility to do this remotely via \gls{vr} solutions to monitor the current construction progress or
to show the current progress to clients \cite{rahimian2020ondemand, vincke2019immersive}. For this, there is the need for measurements
on-site either by combining different sensors \cite{vincke2019immersive} or by using depth images \cite{rahimian2020ondemand}.

Another approach \cite{zaher2018mobile} focuses on monitoring the time and costs. It shows them in an \gls{ar} environment, where the user
sees the building visualized, e.g., on a table as a miniature model.

Applications of the \emph{Task guidance} use case category in the construction phase
\cite{chalhoub2018mixed, dallasega2020bim, riexinger2018mixed, schweigkofler2018development}
help construction workers complete their construction tasks.
This guidance can be done by simply showing the planned model to the worker, so the worker can start replicating it in
the real world \cite{chalhoub2018mixed, dallasega2020bim}. An extension to such approaches is
adding an information system to also show the workers the tasks that need to be done by amending the task
information to the object via \gls{ar} \cite{schweigkofler2018development}. A more enhanced method is
visualizing each task step-by-step with animations and information to guide the worker \cite{riexinger2018mixed}.

Approaches of the \emph{Safety} use case category in the construction phase
\cite{afzal2021evaluating, azhar2017role, hasan2022augmented} assure the safety of stakeholders at the construction site.
This is done by simulating the construction works in a \gls{vr} environment prior to on-site work \cite{afzal2021evaluating, azhar2017role}. Based on this,
safety plans can be created and refined, and potential hazards can be detected and mitigated \cite{afzal2021evaluating, azhar2017role}.
In addition, in case of an accident, these simulations can later be used to
recreate and investigate it \cite{azhar2017role}.

Another approach \cite{hasan2022augmented} is used on-site. Here, a digital twin of a construction crane is created and can be controlled in \gls{ar}
to increase the safety of the operation of the crane.
The approach is only tested with a model of a crane in a laboratory setup. The authors state that the usage of digital twins could be extended
to other construction machinery too.

The last use case category is \emph{Education} in the construction phase \cite{afzal2021evaluating, azhar2017role, saar2019bim}.
It is mainly used by approaches that are also part of the \emph{Safety} use case category to train construction workers to avoid
safety hazards \cite{afzal2021evaluating, azhar2017role}. Another approach argues for using archived \gls{bim} models to train
novices \cite{saar2019bim}.

Some approaches \cite{gomez2019quantitative, herbers2019indoor, huebner2018marker, mahmood2020bim}
focus on \emph{Localization \& Tracking}. These approaches do not focus on concrete use cases but
make certain use cases on-site possible by increasing the precision of localization and tracking on-site.
Here, it is possible to use GPS and device sensors \cite{gomez2019quantitative}, to use one-time marker-based localization
and then to rely on the device sensors \cite{huebner2018marker}, or to match the observed data from the device sensors
with the \gls{bim} model to localize the user \cite{herbers2019indoor, mahmood2020bim}.

\subsubsection{Operation}

Five approaches have been identified in the operation phase \cite{chew2020evaluating, diao2019bim, herbers2019indoor, saar2019bim, xie2020visualized}.
In this phase, most of the approaches focus on \emph{Task guidance}.

\begin{figure}
  \centering
  \subfloat[Navigation \cite{diao2019bim}]{\label{fig:operation:navigation}\includegraphics[width=0.48\textwidth]{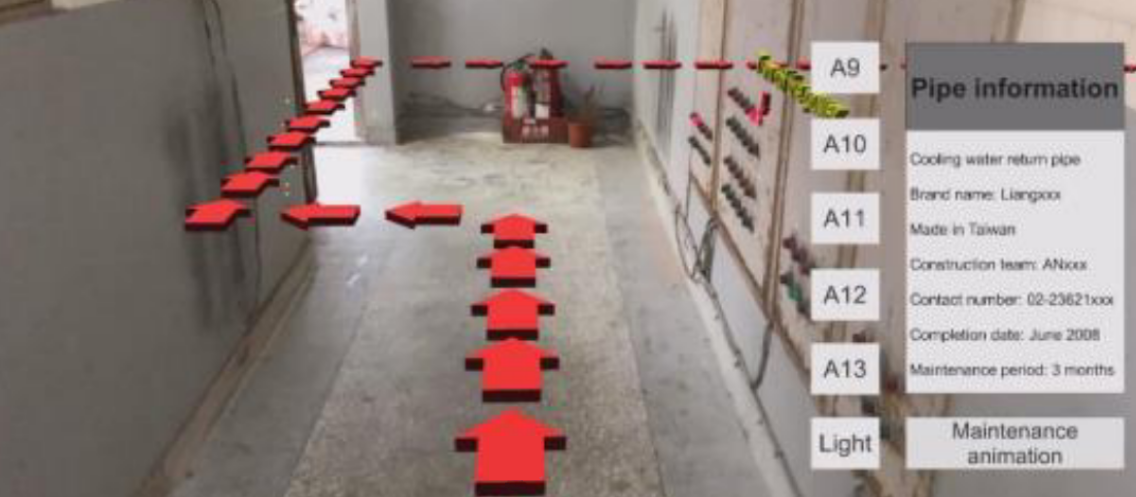}}
  \hfill
  \subfloat[Task guidance \cite{diao2019bim}]{\label{fig:operation:guidance}\includegraphics[width=0.48\textwidth]{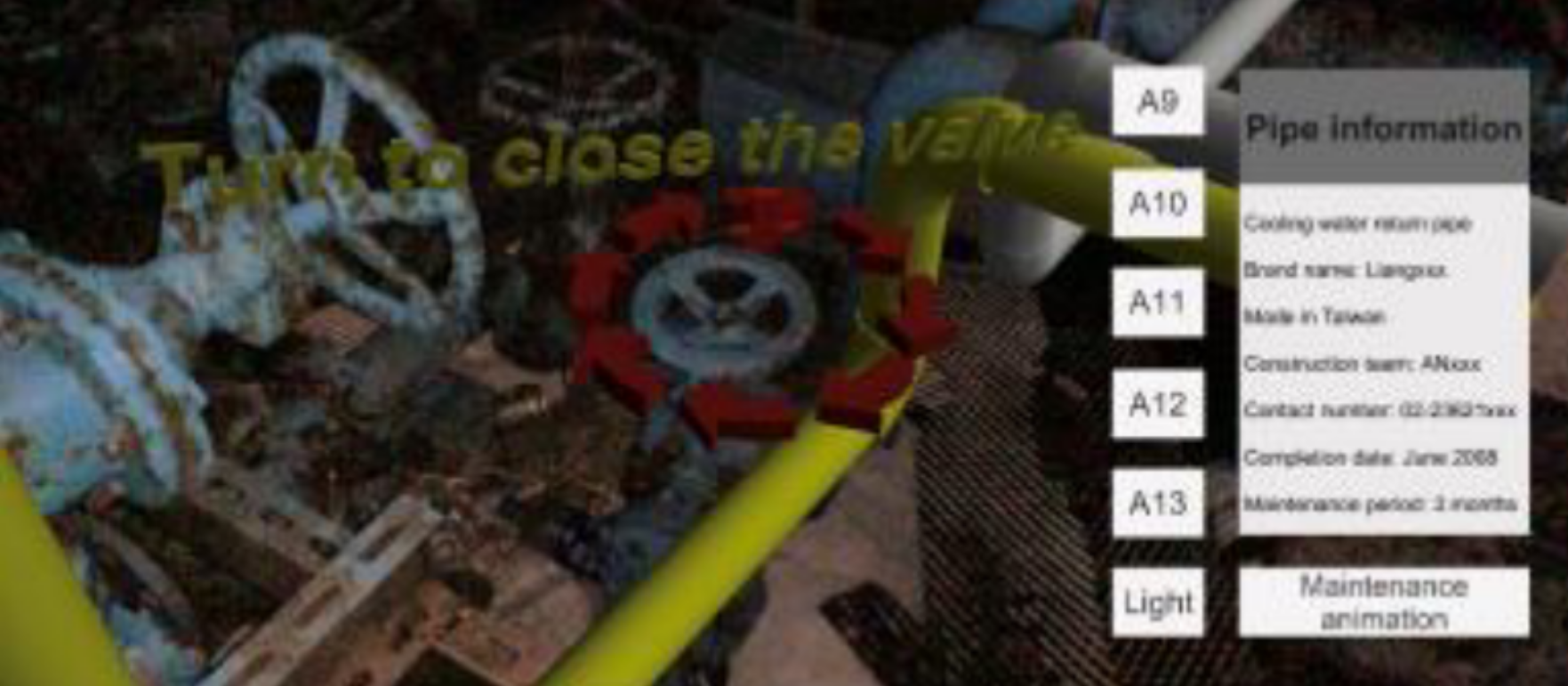}}%
  \caption{Operation-related approaches in the area of AR/VR and BIM}
\end{figure}

Here, \emph{Task guidance} in the maintenance phase \cite{chew2020evaluating, diao2019bim, saar2019bim, xie2020visualized} helps
facility managers or other maintenance staff to complete maintenance tasks. This assistance can start with a system that automatically detects anomalies
via sensors and uses the \gls{bim} model to find the failed asset \cite{xie2020visualized}. Then, \gls{ar} can be used to highlight the failed asset
that might be hidden behind a wall \cite{xie2020visualized, diao2019bim, saar2019bim}. It is also possible to extend this by showing maintenance workers
a path in \gls{ar} that guides them to the failed asset \cite{diao2019bim}. This navigation is shown in \Cref{fig:operation:navigation}, where the red arrows show the
maintenance worker a safe path \cite{diao2019bim}. Some approaches \cite{chew2020evaluating, diao2019bim} also support
maintenance workers by showing them information and animations for a step-by-step guide on how to do their tasks. This guidance is shown in
\Cref{fig:operation:guidance}, where red arrows are used to indicate the direction in which the valve needs to be turned \cite{diao2019bim}.

One approach \cite{diao2019bim} is also part of the \emph{Safety} use case category as it guides the maintenance workers via a safe path to a
failed asset in a dark environment, thus increasing safety. Another approach is part of the \emph{Education} use case category by using
the step-by-step guide for training novice maintenance workers \cite{chew2020evaluating}. This approach is also used to introduce maintenance
workers to facility management basics \cite{chew2020evaluating}. Finally, one of the \emph{Localization \& Tracking} approaches \cite{herbers2019indoor}
also states that it can be used in the \emph{Operation} phase to localize a maintenance worker in the building.

\subsection{Comparison of AR and VR for Different Use Cases}
\label{sec:results:arvr}

In this section, an analysis is done based on whether a use case can be realized with \gls{ar} or \gls{vr}
and the reasons for using \gls{ar} or \gls{vr} for a specific use case. To achieve this, the reasons for selecting \gls{ar} or \gls{vr} are
elicited from the surveyed approaches.

\subsubsection{Design}
In the \emph{Design} phase, three approaches use \gls{ar}
\cite{garbett2021multiuser, khalek2019augmented, schranz2021potentials},
and six use \gls{vr} \cite{kamari2021bim, natephra2017integrating, rahimian2019openbim, vasilevski2020analysing, ventura2020design, zaker2018virtual}.
Thus, \gls{vr} is prevalent in this phase. 

For the \emph{Planning} use case category, one approach uses \gls{ar} \cite{garbett2021multiuser}, and two use \gls{vr} \cite{natephra2017integrating, rahimian2019openbim}.
The \gls{ar} approach combines \gls{ar} with a touch table at which users collaborate \cite{garbett2021multiuser}.
Here, \gls{ar} allows more intuitive collaboration between the users and allows to combine 3D models in \gls{ar} with 2D models shown on the touch table or 2D
printed plans, as the users can still see the real world when using \gls{ar} \cite{garbett2021multiuser}. The first \gls{vr} approach needs to use \gls{vr} because the user should design the indoor lighting situation \cite{natephra2017integrating}. This design task is only possible with \gls{vr} because the user needs
to immerse in the virtual world to experience the different possibilities for lighting in the room \cite{natephra2017integrating}.

The second \gls{vr} approach \cite{rahimian2019openbim} claims to use \gls{ar} and \gls{vr} with a mobile device for \gls{ar} and an \gls{hmd} for \gls{vr}.
Here, it is crucial to notice that the real world is not visible on the mobile device, and only an entirely virtual world is shown.
This \gls{ar} usage is contrary to the definition used in this paper, and thus it is not counted as an \gls{ar} approach in this paper.
The advantage of using \gls{vr} for interior design is that it allows the users to immerse in the environment, and thus make better design decisions \cite{rahimian2019openbim}.
The main disadvantage of \gls{vr} is that it is prone to motion sickness, which needs to be avoided \cite{rahimian2019openbim}.

For the \emph{Review \& Quality Assurance} use case category, two approaches use \gls{ar} \cite{khalek2019augmented, schranz2021potentials},
and three use \gls{vr} \cite{kamari2021bim, ventura2020design, zaker2018virtual}. The first \gls{ar} approach selects \gls{ar} to check a building
for building regulations because it allows showing the planned building on the plot of land on which it is to be built \cite{schranz2021potentials}. This visualization
allows the users to check whether it fits the neighboring buildings \cite{schranz2021potentials}. Additionally, when multiple people discuss the planned building,
an \gls{ar} approach allows for better collaboration \cite{schranz2021potentials}. Here, \gls{ar} allows showing the planned house as a model on a table together with the
neighboring buildings, and the discussants sit around the table all equipped with an \gls{ar} device \cite{schranz2021potentials}.

Another approach, which allows a user to compare different façade designs, selects \gls{vr} as a medium because a user is more immersed in it \cite{kamari2021bim}.
The immersion is also mentioned by Zaker and Coloma \cite{zaker2018virtual}, who highlight the advantage that \gls{vr} allows users to see the actual dimensions of the
building. All of these approaches \cite{schranz2021potentials, kamari2021bim, zaker2018virtual} emphasize that novices that do not understand complicated 2D plans,
like clients and owners, profit from the usage of \gls{ar}/\gls{vr} in the review process. Also, in this use case, the disadvantage of the \gls{vr} approaches is the potential for motion
sickness \cite{ventura2020design, zaker2018virtual}. Here, Ventura et al. \cite{ventura2020design} mention
that novice users should not be trained before the actual review in \gls{vr} to avoid prior motion sickness. Additionally, they add breaks to their
design review protocol to allow users to recover from potential motion sickness.

One of the two approaches that review the design’s maintainability uses \gls{ar} \cite{khalek2019augmented}, and the other one uses \gls{vr} \cite{zaker2018virtual}.
The \gls{vr} approach reports that joysticks, which are used by many \gls{vr} devices today, are not that easy to use for the simulation of the maintenance task,
and participants needed some training \cite{zaker2018virtual}. In this context, the authors report that they selected \gls{ar} because
users can see their own bodies and can do the same tasks a facility manager would need to do \cite{khalek2019augmented}. This also enables novices to find design flaws
that lead to poor maintainability \cite{khalek2019augmented}.

For the \emph{Education} use case category, only a \gls{vr} approach is used \cite{vasilevski2020analysing}. Here, a significant aspect is an ability to change
the time of day to see the design in different lighting situations, which is only possible to experience in \gls{vr}. Still, also in this approach, motion sickness
is reported by a few of the participants \cite{vasilevski2020analysing}.

\subsubsection{Construction}
In the construction phase, 17 approaches use \gls{ar}
\cite{chalhoub2021augmented, chalhoub2018mixed, dallasega2020bim, garbett2021multiuser, gomez2019quantitative,
hasan2022augmented, herbers2019indoor, hernandez2018lean, huebner2018marker, lou2017study, mahmood2020bim,
mirshokraei2019web, ratajczak2019bim, riexinger2018mixed, saar2019bim, schweigkofler2018development, zaher2018mobile},
and five approaches use \gls{vr} \cite{afzal2021evaluating, azhar2017role, dallasega2020bim, rahimian2020ondemand, vincke2019immersive}.
Thus, \gls{ar} is prevalent in this building's lifecycle phase. 

In the \emph{Planning} use case category, one approach uses \gls{ar} \cite{garbett2021multiuser}, and one uses \gls{vr} \cite{dallasega2020bim}.
The participants in the \gls{vr} approach reported that the \gls{vr} environment is too complicated without prior training \cite{dallasega2020bim}.
The \gls{ar} approach \cite{garbett2021multiuser} is the same approach used for planning in the design phase. Thus, the same reasons for the selection of
\gls{ar} apply, which are intuitive collaboration, combination with other 2D screens, and the possibility to include analog artifacts \cite{garbett2021multiuser}.

Nine approaches
\cite{chalhoub2021augmented, lou2017study, mirshokraei2019web, ratajczak2019bim, riexinger2018mixed, saar2019bim, schweigkofler2018development, zaher2018mobile, hernandez2018lean}
use \gls{ar}, and two use \gls{vr} \cite{rahimian2020ondemand, vincke2019immersive} in the
\emph{Review \& Quality Assurance} use case category. Thus, \gls{ar} is prevalent in this use case category in the construction phase.
Here, all \gls{ar} approaches can be used on-site, which is not
possible with immersive \gls{vr} as users must see the real world on-site. One approach also mentions the advantage for site managers to be able to
walk through the building and see the physical construction objects amended with the necessary information \cite{ratajczak2019bim}. Additionally, it is only
possible with \gls{ar} to compare the as-planned and as-built state on-site by amending the real world with the as-planned objects to see deviations \cite{chalhoub2021augmented}.
Still, the issue is that only large deviations or missing parts are easily detectable via \gls{ar} \cite{chalhoub2021augmented}.

Additionally, the tracking of \gls{ar}
is imperfect, and thus, the model can have an offset or starts to drift \cite{chalhoub2021augmented, mirshokraei2019web, ratajczak2019bim}. This imperfection even leads to the
fact that measurements are imprecise if done in \gls{ar} \cite{mirshokraei2019web}. Furthermore, there is the issue of
occlusions, that some objects are displayed in front of real-world objects although they should be displayed behind \cite{mirshokraei2019web}. This occlusion is even safety critical
for on-site usage as the user's vision could be hindered \cite{mirshokraei2019web}.

Another possibility is to do the quality assurance and deficit detection remotely in an office with \gls{vr} \cite{rahimian2020ondemand, vincke2019immersive}.
Here, the advantages are the possibility to easily show the progress to the client without going to the construction site, and the construction works
are not interrupted \cite{rahimian2020ondemand}. Additionally, it does not rely on the imprecise tracking and localization of \gls{ar} and can
provide more accurate results \cite{rahimian2020ondemand, vincke2019immersive}. These accurate results allow for more fine-grained quality control \cite{vincke2019immersive}.
The disadvantages of this approach are that the monitoring does not happen in real-time, and sensors must be used beforehand to get a representation of
the current on-site status \cite{rahimian2020ondemand, vincke2019immersive}.

One approach \cite{zaher2018mobile} uses \gls{ar} in another way. Here, instead of overlaying the real world with the as-planned state, a
virtual model of the building is presented in \gls{ar}, for example, standing on a table. For this approach, it is also necessary to acquire the
current state on-site beforehand \cite{zaher2018mobile}. However, compared to the \gls{vr} applications, this approach can also be
used on-site, as the user is not immersed in the model \cite{zaher2018mobile}. On the other side, the approach only allows for an overview of the progress of the construction works and not
detailed deficit detection \cite{zaher2018mobile}.

All approaches \cite{chalhoub2018mixed, dallasega2020bim, riexinger2018mixed, saar2019bim, schweigkofler2018development} of the \emph{Task guidance} use case category use \gls{ar}.
As already mentioned in the previous sections, the usage of \gls{vr} on-site is not possible, and thus, as \emph{Task guidance} has to be done on-site, the usage of \gls{ar} is required.
Additionally, \gls{ar} is felt intuitive by users for \emph{Task guidance} \cite{dallasega2020bim}. Here, already significant performance improvements of workers were shown in
laboratory setups \cite{chalhoub2018mixed, dallasega2020bim}. To provide the user with visualized task guidance, a 3D model of the to-be-constructed part of
the building must exist, and thus the \gls{bim} model must contain these details \cite{chalhoub2018mixed}.

In the \emph{Safety} use case category, one approach \cite{hasan2022augmented} uses \gls{ar}, and the other two \cite{afzal2021evaluating, azhar2017role} use \gls{vr}.
The \gls{ar} approach supports the workers on-site to enhance their safety there \cite{hasan2022augmented}. The \gls{vr} approaches are used off-site to plan and review safety
\cite{afzal2021evaluating, azhar2017role}. Here, the communication of safety to the construction workers is increased, as they can immerse in the on-site situation without
being there \cite{afzal2021evaluating}. Through this, they are able to get to know the on-site environment and experience safety-critical situations without danger to their lives
\cite{afzal2021evaluating}. The problems are the need for precise \gls{bim} models that include not only the building but also the different construction stages with the
machines and tools used there \cite{afzal2021evaluating}. This modeling task can lead to increased costs \cite{azhar2017role}. Additionally, users of the system can suffer from
motion sickness \cite{azhar2017role}.

In the \emph{Education} use case category, one approach uses \gls{ar} \cite{saar2019bim}, and two use \gls{vr} \cite{afzal2021evaluating, azhar2017role}.
The \gls{ar} approach \cite{saar2019bim} only states to use \gls{ar}. Here, the prototype presented does not use \gls{ar} yet but should include the real world in newer versions, and thus
we count the approach as \gls{ar}. They do not mention specific reasons why \gls{ar} should be used instead of \gls{vr} for education. The two \gls{vr} approaches
\cite{afzal2021evaluating, azhar2017role}
are the two mentioned in the \emph{Safety} use case category. They also use the \gls{vr} system to train workers, and thus the same reasons mentioned in the last paragraph apply.

\subsubsection{Operation}
In the operation phase, four approaches use \gls{ar} \cite{diao2019bim, herbers2019indoor, saar2019bim, xie2020visualized}, and one
uses both \gls{ar} and \gls{vr} for different use cases \cite{chew2020evaluating}. Thus, \gls{ar} is prevalent in this phase.

All approaches \cite{chew2020evaluating, diao2019bim, saar2019bim, xie2020visualized} of the \emph{Task guidance} use case category in the operation phase use \gls{ar}.
Obviously, \gls{vr} is not possible in this phase, as the maintenance workers must do their tasks on-site and need to do the tasks in the real world. \Gls{ar} supports
them in finding the failed asset by guiding them toward it with a path shown in \gls{ar} \cite{diao2019bim}. Here, this is also part of the \emph{Safety} use case category
as it improves the safety of the maintenance workers to find a safe path in a dark environment \cite{diao2019bim}.
Additionally, \gls{ar} allows highlighting the asset, which might be hidden behind a wall
\cite{diao2019bim, xie2020visualized}, and can give them instructions on how to fix the failed asset by highlighting different parts of it and showing animations
of what they need to do \cite{diao2019bim, chew2020evaluating}. Again, the potential model drift of \gls{ar} systems is a problem \cite{diao2019bim}.

One approach \cite{chew2020evaluating} uses \gls{ar} and \gls{vr} for the \emph{Education} use case category. Here, the approach uses \gls{ar} to support
novices while doing maintenance work \cite{chew2020evaluating}. Additionally, \gls{vr} is used to teach them the basics of facility management in an
immersive environment \cite{chew2020evaluating}. Here, the advantage is that the participants do not need to be in the same physical location and
can still experience an interactive 3D world to learn together \cite{chew2020evaluating}.

\subsection{Discussion}
\label{sec:results:discussion}
In this section, the results are concluded to answer the research questions.

The first research question deals with the use cases in which \gls{ar} and \gls{vr} are used together with \gls{bim}.
Here, the five use case categories \emph{Planning}, \emph{Review \& Quality Assurance}, \emph{Task guidance}, \emph{Safety}, and
\emph{Education} are defined. Additionally, some papers deal with the technical detail of on-site localization and tracking. The results show that
the most supported use case category is \emph{Review \& Quality Assurance}, especially in the \emph{Construction} phase, followed by
\emph{Task guidance} in the \emph{Construction} and \emph{Operation} phases. Only a few papers focus on the other three use case categories.

The second research question deals with the reasons why \gls{ar} or \gls{vr} should be used for a use case.
In the \emph{Design} phase, the results show that \gls{ar} and \gls{vr} support novices that do not understand complicated 2D plans.
\Gls{ar} is used in this phase for rough planning and collaborative tasks, for tasks that involve visualizing the building in the real world, or
for tasks that require the simulation of physical work.
\Gls{vr} is used for detailed planning tasks, where a user benefits from the immersive environment.

In the \emph{Construction} phase, the usage of \gls{ar} and \gls{vr} depends more on the use case. Here, for the \emph{Planning} use case category, the
same arguments as for the \emph{Design} phase are valid. For all use cases that need to be done on-site, the usage of \gls{ar} is mandatory, as it
is impossible to use \gls{vr} on-site. Thus, for \emph{Task guidance}, only \gls{ar} is used. For \emph{Review \& Quality Assurance}, \gls{ar} is often
used to allow for on-site usage, but it is also possible to use \gls{vr}, which allows for more precise results but also needs prior measurements.
In the \emph{Safety} use case category, \gls{vr} is used for
planning steps, and \gls{ar} is used to support the safety on-site. For \emph{Education}, \gls{ar} or \gls{vr} might be used.

In the \emph{Operation} phase, for all use case categories, \gls{ar} is used, because maintenance is a task that needs to be done
on-site, and thus only \gls{ar} is possible. The only exception is the \emph{Education} use case category. Here, \gls{ar} is
still used if novices are trained on-site, but \gls{vr} can be used for off-site education.

\section{Further Research}
\label{sec:furtherResearch}
Most approaches focus on the \emph{Construction} phase, second most on the \emph{Design} phase. This fits the findings of earlier surveys that
also see the construction and design phase as the dominant research subject \cite{sidani2021tools, wang2013augmented}. Thus, one open
issue is to have more studies in the \emph{Operation} phase. Additionally, most of the approaches in the \emph{Operation} phase
focus on maintenance tasks. Only one uses sensors that are placed in the building to monitor the building. Here, more research is possible in the
domain of connecting smart \gls{iot} devices, \gls{ar}/\gls{vr}, and \gls{bim} not only for maintenance 
but also for the operation of a building, like controlling light and heating systems.

None of the surveyed approaches does a direct comparison between \gls{ar} and \gls{vr} solutions. Especially for use cases where both technologies could be
used, e.g., in the \emph{Design} phase and \emph{Review \& Quality Assurance} in the \emph{Construction} phase,
this would be beneficial to get a deeper insight into the usability of such approaches in direct comparison.

Only a few approaches focus on the \emph{Safety} use case category. Especially in the \emph{Construction} phase, only one approach is used directly on-site
to increase safety. Here, more research in this area would be beneficial to bring the simulation features that \gls{bim} provides for the
\emph{Construction} phase directly on-site. Furthermore, in the \emph{Operation} phase, only one approach increases safety. Also, here,
more research is possible. Additionally, research is needed on how \gls{ar} devices can be used safely on-site.

Many of the surveyed papers focus on the \emph{Review \& Quality Assurance} use case category. Here, a shift towards allowing workers to use such approaches
and directly supporting them with \emph{Task guidance} would be beneficial. Through this, errors could be avoided and found in the early stages and easily resolved.

From the classified literature, only two approaches deal with the \emph{Planning} use case category in the \emph{Construction} phase. Here, more research
to leverage the simulation features of \gls{bim} for detailed planning of the on-site environment is required. Currently, \gls{bim} is only used
to look at the model or add some information to objects in the \gls{bim} model. Additionally, this planned model of the on-site environment could
then be used on-site to place materials, machines, and other equipment in the correct position on-site.

\section{Limitations}
\label{sec:limitations}

First of all, there are limitations to the literature selection progress.
Here, papers could be missed through the selected literature database and search terms. Additionally, only open-access papers and papers
with a citation count of ten or more citations were examined. This could affect the results in three aspects. First, the given statistics
about the lifecycle phases, use cases, used technology, devices, and combination of these categories could differ. Second, use case categories could be
missing due to missing papers. 

Still, the statement that there is missing research should be valid. If an approach is not in the surveyed literature, either it is not open-access, or it is not cited often enough.
Thus, if a topic is already researched in detail, it is unlikely that all papers in that research field are not open access. Additionally, if there is much research in
a field, the papers of the field cite each other leading to many citations on some papers, which should then be included in this survey.

\section{Conclusion}
\label{sec:conclusion}
While an increasing number of studies in HCI, construction, or engineering have shown the potential of using AR and VR technology together with BIM, often research remains focused on individual explorations and key design strategies. In addition to that, a systematic overview and discussion of recent approaches are missing. Therefore, we have systematically reviewed recent approaches combining \gls{ar}/\gls{vr} with \gls{bim} and categorized the literature by the building's lifecycle phase while systematically describing relevant use cases.

Future work should examine approaches of specific lifecycle phases or use case categories in more detail. Additionally, it would be helpful to analyze the surveyed literature based on other categories. Here, especially
aspects such as usability, UX as well as safety, and security could be investigated in more detail to provide further insights about the applicability of AR/VR and BIM in the \gls{aec-fm} industry.

%
%
%
\bibliographystyle{splncs04}
\bibliography{bibliography}
\end{document}